\documentclass[preprint,aps,prd,superscriptaddress]{revtex4}

\RequirePackage{xspace}

\usepackage{epsfig}
\usepackage{graphicx}
\usepackage{color}

\def\sss{\scriptsize}
\def\nn{\nonumber}

\def\dA{\delta^A_\lambda}
\def\dP{\delta^P_\lambda}
\def\dT{\delta^T_\lambda}
\def\dC{\delta^C_\lambda}
\def\dpewc{\delta^{CEW}_\lambda}
\def\dpew{\delta^{EW}_\lambda}

\def\beq{\begin{equation}}
\def\eeq{\end{equation}}
\def\bea{\begin{eqnarray}}
\def\eea{\end{eqnarray}}

\def\nn{\nonumber}
\def\sss{\scriptscriptstyle}

\def\barp{{\raise.35ex\hbox{${\sss (}$}}---{\raise.35ex\hbox{${\sss )}$}}}
\def\bdbarp{\hbox{$B_d$\kern-1.4em\raise1.4ex\hbox{\barp}}}
\def\bsbarp{\hbox{$B_s$\kern-1.4em\raise1.4ex\hbox{\barp}}}

\def\roughly#1{\mathrel{\raise.3ex\hbox{$#1$\kern-.75em\lower1ex\hbox{$\sim$}}}}
\def\barpk{{\raise.35ex\hbox{${\sss (}$}}--{\raise.35ex\hbox{${\sss )}$}}}
\def\bbarp{\hbox{$B$\kern-0.9em\raise1.4ex\hbox{\barpk}}}

\def\dA{\delta^A_\lambda}
\def\dP{\delta^P_\lambda}
\def\dT{\delta^T_\lambda}
\def\dC{\delta^C_\lambda}
\def\dpewc{\delta^{CEW}_\lambda}
\def\dpew{\delta^{EW}_\lambda}

\def\nn{\nonumber}

\def\adir00{{a_{\sss dir}^{00}}}

\def\B00{B^{00}}
\def\Bp0{B^{+0}}

\def\dsp{\displaystyle}

\newcommand{\tev}{\ensuremath{\mathrm{Te\kern -0.1em V}}\xspace}
\newcommand{\gev}{\ensuremath{\mathrm{Ge\kern -0.1em V}}\xspace}
\newcommand{\mev}{\ensuremath{\mathrm{Me\kern -0.1em V}}\xspace}
\newcommand{\kev}{\ensuremath{\mathrm{ke\kern -0.1em V}}\xspace}
\newcommand{\ev}{\ensuremath{\mathrm{e\kern -0.1em V}}\xspace}
\newcommand{\gevc}{\ensuremath{{\mathrm{Ge\kern -0.1em V\!/}c}}\xspace}
\newcommand{\mevc}{\ensuremath{{\mathrm{Me\kern -0.1em V\!/}c}}\xspace}
\newcommand{\gevcc}{\ensuremath{{\mathrm{Ge\kern -0.1em V\!/}c^2}}\xspace}
\newcommand{\mevcc}{\ensuremath{{\mathrm{Me\kern -0.1em V\!/}c^2}}\xspace}

\begin{document}

\preprint{IMSc/2006/12/27}

\title{Generalized analysis on $B\to K^* \rho$ \\
within and beyond the Standard Model \\
$-$ Can it help understand the $B \to K \pi$ puzzle?}
\author{C.~S.~Kim}\email{cskim@yonsei.ac.kr} \affiliation{Department
  of Physics, Yonsei University, Seoul 120-479, Korea}
\author{Sechul~Oh}\email{scoh@phya.yonsei.ac.kr} \affiliation{Department
  of Physics, Yonsei University, Seoul 120-479, Korea}
\author{Chandradew Sharma}\email{sharma@imsc.res.in}
\affiliation{The Institute of
  Mathematical Sciences, Taramani, Chennai 600113, India}
\author{Rahul Sinha}\email{sinha@imsc.res.in}\affiliation{The Institute of
  Mathematical Sciences, Taramani, Chennai 600113, India}
\author{Yeo Woong Yoon}\email{ywyoon@yonsei.ac.kr}\affiliation{Department
  of Physics, Yonsei University, Seoul 120-479, Korea}

\date{\today}
%
%
\begin{abstract}
  \noindent
  We study $B \to K^* \rho$ modes that are analogues of the much
  studied $B\to K \pi$ modes with $B$ decaying to two vector mesons
  instead of pseudoscalar mesons, using topological amplitudes in the
  quark diagram approach.  We show how $B \to K^*\rho$ modes can be
  used to obtain many more observables than those for $B \to K \pi$
  modes, even though the quark level subprocesses of both modes are
  exactly the same.  All the theoretical parameters (except for the
  weak phase $\gamma$), such as the magnitudes of the topological
  amplitudes and their strong phases, can be determined in terms of
  the observables without any model-dependent assumption.  We
  demonstrate how $B\to K^*\rho$ can also be used to verify if there
  exist any relations between theoretical parameters, such as the
  hierarchy relations between the topological amplitudes and possible
  relations between the strong phases.  Conversely, if there exist
  reliable theoretical estimates of amplitudes and strong phases, the
  presence of New physics could be probed.  We show that if the tree and
  color-supressed tree are related to the electroweak penguins and
  color-supressed electroweak penguins, it is not only possible  to
  verify the validity of such relations but also to have a clean
  measurement of New Physics parameters. We also present a numerical
  study to examine which of the observables are more sensitive to New
  Physics.
\end{abstract}

\maketitle

\section{Introduction}
\label{sec:1}

The study of two-body hadronic $B$ decays provides good opportunities
to test the Standard Model (SM) and further to probe possible new
physics (NP) effects beyond the SM.  Large numbers of B mesons have
been produced at the B factories enabling accurate measurements of
branching ratios and direct CP asymmetry for many modes. The $B \to
VV$ modes, where $V$ denotes a vector meson, have the advantage that
they provide many more observables, compared with those being measured
in $B \to PP$ ($e.g.$, $B \to K \pi$) or $B \to VP$ ($e.g.$, $B \to
K^* \pi$) modes, where $P$ denotes a pseudoscalar meson, due to spins
of the final state vector mesons. Since the first observation of $B
\to K^* \phi$ by CLEO Collaboration~\cite{Bergfeld:1998ik}, several
$B$ decays to two charmless vector mesons, such as $B \to K^* \rho$
and $B \to \rho \rho$, have been reported by BABAR and BELLE
Collaboration~\cite{HFAG,Yao:2006px,Aubert:2003mm,Aubert:2004xc,
Aubert:2006ae,Abe:2004mq}.
In fact, polarization measurements for several such modes have already
been reported.

Recent experimental results~\cite{HFAG,Yao:2006px,Abe:2005fz}
for the $B \to K \pi$ mode, show deviations from SM expectations; the
discrepancy, commonly being referred to as the ``$B \to K\pi$
puzzle.''  The dominant quark level subprocesses for $B \to K \pi$
decays are $b \to s \bar q q$ ($q = u,d$) penguin processes which are
potentially sensitive to NP effects.  Many efforts have been made to
resolve the puzzle~~\cite{Mishima:2004um,Kim:2005jp}.
A model-independent study shows that the experimental
data strongly indicate large enhancements of both the electroweak (EW)
penguin and the color-suppressed tree contributions~\cite{Kim:2005jp}.
The $B\to K\pi$ modes have certain inherent limitations. The four
$B\to K\pi$ decay modes can experimentally yield at most 9
observables: four each of the branching ratios and direct CP
asymmetries and one time-dependent CP asymmetry. Clearly, the 9
observables are insufficient to determine all the 12 theoretical
parameters~\cite{Kim:2005jp} needed to describe these decay modes.
One hence needs to make some assumptions.  Traditionally, assumptions
have often been made on sizes of the topological amplitudes as well
as on the strong phases of the different topologies.

The $B \to K^* \rho$ modes are the $B \to VV$ analogues of $B \to K
\pi$ modes, in the sense that the quark level processes of both modes
are exactly the same.  Thus, it is expected that if there appear any
NP effects through $B \to K \pi$, then similar NP effects will appear
through $B \to K^* \rho$ as well. However, the study of $B\to VV$
modes necessitates performing an angular analysis in order to obtain
the helicity amplitudes. While angular analysis is often regarded as
an additional complication needed due to the presence of both CP-even
and CP-odd components that dilute the time dependent CP asymmetry, it
can provide an impressive gain in terms of the large number of
observables.  We note that preliminary polarization measurement for
$B^+\to K^{*+} \rho^0$, $B^+\to K^{*0} \rho^+$ and $B^0\to K^{*0}
\rho^0$ have already been done~\cite{HFAG,Yao:2006px,Aubert:2003mm,Abe:2004mq}.

In this work, we study $B \to K^* \rho$ decays in a model-independent
approach. We will show that in comparison to the $B \to K \pi$ modes
that yield 9 observables, the $B \to K^* \rho$ modes result in a total
of 35 independent observables.  A theoretical description of $B\to
K^*\rho$, however, requires 36 independent parameters, which is still
one short of the number of possible observables.  While the large
number of observables may not seem like a distinct advantage at first,
we will argue that they provide valuable insights into resolving the
``$B\to K\pi$ puzzle.''  Our goal is two-fold.  First, we try to
determine all the relevant theoretical parameters describing the decay
amplitudes of $B \to K^* \rho$ in a model-independent way in terms of
experimental observables.  These determined theoretical parameters can
be compared with the corresponding model estimates.  The information
will be very useful for improving model calculations, such as those
based on QCD factorization~\cite{Beneke:2003zv}, perturbative
QCD~\cite{Keum:2000wi}, and so on.  Second, we try to suggest certain
tests of the SM that may reveal NP effects if they appear in $B \to
K^* \rho$ decays.  Any indication of NP effects in $B \to K^* \rho$
will provide valuable hints on possible NP effects in $B \to K \pi$
decays.  For our goal, the decay amplitudes of $B \to K^* \rho$ are
decomposed into linear combinations of the topological amplitudes in
the quark diagram approach~\cite{Gronau:1994rj}. We then
focus on how to extract all the theoretical parameters, including the
magnitudes of the topological amplitudes and their strong phases, in
terms of experimental observables. As it turns out, all the parameters
can be determined in analytic forms.  We also propose tests of
conventional hierarchy relations between the topological amplitudes
and of possible relations between the relevant strong phases within
the SM.  A breakdown of these relations may indicate possible NP
contributions appearing in $B \to K^* \rho$ decays, as well as in the
analogous mode $B\to K\pi$. One could hence verify if NP is the source
of the ``$B\to K\pi$ puzzle.''

The paper is organized as follows.  A general formalism for $B \to K^*
\rho$ is presented in Sec.~\ref{sec:2}.  In Sec.~\ref{sec:3} we
explicitly show that it is possible to obtain analytic solutions to
all the theoretical parameters in terms of observables.  In
Sec.~\ref{sec:4} we discuss how to examine the conventional hierarchy
of the topological amplitudes and possible relations of their strong
phases.  We conclude in Sec.~\ref{sec:5}.

\section{Formalism for $B\to K^*\rho$ decays}
\label{sec:2}

The decay amplitudes for four $B \to K^* \rho$ modes can be written in terms of
the topological amplitudes in the quark diagram approach as
\begin{eqnarray}
\label{decayamp-a1}
&& A^{0+}_{\lambda} \equiv A_{\lambda}(B^+ \to K^{*0} \rho^+)
 = V_{ub}^* V_{us} A^{\prime}_{\lambda} +V_{tb}^* V_{ts} P^{\prime}_{\lambda} , \\
\label{decayamp-a2}
&& A^{+0}_{\lambda} \equiv A_{\lambda}(B^+ \to K^{*+} \rho^0)
 = - {1 \over \sqrt{2}}
  \left[ V_{ub}^* V_{us} (T^{\prime}_{\lambda} +C^{\prime}_{\lambda}
   +A^{\prime}_{\lambda})
  +V_{tb}^* V_{ts} (P^{\prime}_{\lambda} +P^{EW \prime}_{\lambda}
   +P^{EW \prime}_{C, \lambda}) \right], \\
\label{decayamp-a3}
&& A^{+-}_{\lambda} \equiv A_{\lambda}(B^0 \to K^{*+} \rho^-)
 = - \left[ V_{ub}^* V_{us} T^{\prime}_{\lambda}
  +V_{tb}^* V_{ts} (P^{\prime}_{\lambda} +P^{EW \prime}_{C, \lambda}) \right], \\
\label{decayamp-a4}
&& A^{00}_{\lambda} \equiv A_{\lambda}(B^0 \to K^{*0} \rho^0)
 = - {1 \over \sqrt{2}} \left[ V_{ub}^* V_{us} C^{\prime}_{\lambda}
  -V_{tb}^* V_{ts} (P^{\prime}_{\lambda} -P^{EW \prime}_{\lambda}) \right],
\label{decayamp}
\end{eqnarray}
where $V_{ij} ~(i=u, t; ~ j=s, b)$ are Cabibbo-Kobayashi-Maskawa (CKM) matrix elements
and the subscript $\lambda=\{0,\|,\perp\}$ denotes the helicity of the amplitudes.
The amplitudes $T^{\prime}$, $C^{\prime}$, $A^{\prime}$, $P^{\prime}$,
$P^{EW \prime}$, and $P^{EW \prime}_{C}$ are defined as
\begin{eqnarray}
&& T^{\prime} \equiv T +P_{uc} +E_{uc} ~,
\label{tildeT} \\
&& C^{\prime} \equiv C -P_{uc} -E_{uc} ~,
\label{tildeC} \\
&& A^{\prime} \equiv A +P_{uc} +E_{uc} ~,
\label{tildeA} \\
&& P^{\prime} \equiv P_{tc} +E_{tc} -{1 \over 3} P^{EW}_C
 +{2 \over 3} E^{EW}_C ~,
\label{tildeP} \\
&& P^{EW \prime} \equiv P^{EW} +E^{EW}_C ~,
\label{tildePEW} \\
&& P^{EW \prime}_{C} \equiv P^{EW}_C - E^{EW}_C ~,
\label{tildePEWC}
\end{eqnarray}
where $P_{ic} \equiv P_i - P_c$ and $E_{ic} \equiv E_i - E_c$ $(i = u,t)$.
The topological amplitude $T$ is a color-favored tree amplitude, $C$ is a
color-suppressed tree, $A$ is an annihilation, $P_j~(j = u,c,t)$ is a QCD penguin,
$E_j$ is a penguin exchange, $P^{EW}$ is a color-favored electroweak (EW) penguin,
$P^{EW}_C$ is a color-suppressed EW penguin, $E^{EW}_C$ is a color-suppressed
EW penguin exchange diagram.
We follow and generalize the notation used in Ref.~\cite{Kim:2005jp}.

The relative sizes among these topological amplitudes are roughly estimated
\cite{Gronau:1994rj} as
\begin{eqnarray}
\label{su3est}
1 &:& |V_{tb}^* V_{ts}~ P_{tc}|, \nn \\
\mathcal{O}(\bar \lambda) &:& |V_{ub}^* V_{us}~ T|,~ |V_{tb}^* V_{ts}~ P^{EW}|, \nn \\
\mathcal{O}(\bar \lambda^2) &:& |V_{ub}^* V_{us}~ C|,~ |V_{tb}^* V_{ts}~ P^{EW}_C|, \nn \\
\mathcal{O}(\bar \lambda^3) &:& |V_{ub}^* V_{us}~ A|,~ |V_{ub}^* V_{us}~ P_{uc}|,
\end{eqnarray}
where $\bar \lambda \sim 0.2$.  For the relative size of $|V_{ub}^* V_{us}~ P_{uc}|$, one
can roughly estimate that
\begin{equation}
\label{puc-size}
\left| \frac{V^*_{ub}V_{us}~ P_{uc}}{V^*_{tb}V_{ts}~ P_{tc}} \right|
\sim {\bar \lambda}^2 \left|\frac{P_{uc}}{P_{tc}}\right|.
\end{equation}
Note that $| P_u |$ and $| P_c |$ are smaller than $| P_t |$~\cite{Baek:2005tj}, and
more precisely it can be estimated that $0.2 < | P_{uc} / P_{tc} |< 0.4$
within the perturbative calculation~\cite{Buras:1994pb}.  Therefore, we assume
$| (V^*_{ub}V_{us}~ P_{uc}) / (V^*_{tb}V_{ts}~ P_{tc}) |
\sim \mathcal{O}(\bar \lambda^3)$ for our analysis.

Now we re-express Eqs.~(\ref{decayamp-a1})$-$(\ref{decayamp-a4}) as
\begin{eqnarray}
\label{decayamp-1}
\mathcal{A}^{0+}_\lambda
 &=& e^{i\gamma}\tilde A_\lambda e^{i\dA}
         - \tilde P_\lambda e^{i\dP}~, \\
\label{decayamp-2}
\mathcal{A}^{+0}_\lambda
 &=& - {1 \over \sqrt{2}}
  \Big[e^{i\gamma}(\tilde T_\lambda e^{i\dT} +\tilde C_\lambda e^{i\dC}
  +\tilde A_\lambda e^{i\dA}) \nn\\
 &\mbox{}&
  - (\tilde P_\lambda e^{i\dP}+\tilde P^{EW}_\lambda e^{i\dpew}
  +\tilde P^{EW}_{C,\lambda} e^{i\dpewc} ) \Big]~, \\
\label{decayamp-3}
\mathcal{A}^{+-}_\lambda
 &=& - \Big[e^{i\gamma}\tilde T_\lambda e^{i\dT}
   - (\tilde P_\lambda e^{i\dP}+\tilde P^{EW}_{C,\lambda} e^{i\dpewc} )\Big]~, \\
\label{decayamp-4}
\mathcal{A}^{00}_\lambda
 &=& - {1 \over \sqrt{2}} \left[e^{i\gamma}\tilde C_\lambda e^{i\dC}
    + (\tilde P_\lambda e^{i\dP}-\tilde P^{EW}_\lambda e^{i\dpew}) \right]~,
\end{eqnarray}
where the $\gamma$ and $\delta_{\lambda}$'s are the weak phase and the relevant strong
phases, respectively.
We note that isospin symmetry relates the amplitudes for these 4 decay modes and their
conjugate modes by the relations:
\begin{eqnarray}
\frac{1}{\sqrt{2}}\Big(\mathcal{A}_{\lambda}^{0+} - \mathcal{A}_{\lambda}^{+-}\Big)
 =\mathcal{A}_{\lambda}^{00} - \mathcal{A}_{\lambda}^{+0}~,
\label{eq:isospin-1} \\
\frac{1}{\sqrt{2}}\Big(\bar{\mathcal{A}}_{\lambda}^{0+}
 - \bar{\mathcal{A}}_{\lambda}^{+-}\Big)
 =\bar{\mathcal{A}}_{\lambda}^{00} - \bar{\mathcal{A}}_{\lambda}^{+0}~.
\label{isospin1}
\end{eqnarray}

It should be emphasized that as mentioned in
Ref.~\cite{Imbeault:2006nx}, for the $B\to K\pi$ mode, the above
expressions in Eqs.~(\ref{decayamp-1})$-$(\ref{decayamp-4}) for the
decay amplitudes describe not only the SM contributions but also {\em any}
possible NP effects that contribute to the amplitude.  Consider for
instance the contribution of NP with an amplitude $N_\lambda
e^{i\delta_\lambda} e^{i\phi_{NP}}$. This amplitude may be re-expressed
using reparametrization invariance~\cite{Botella:2005ks} as a sum of
two contributions with one term having no weak phase and the other term
having a weak phase $\gamma$, {\it i.e.} $N_\lambda
e^{i\delta_\lambda} e^{i\phi_{NP}}\equiv N_1^\lambda e^{i\delta_\lambda}+N_2^\lambda
e^{i\delta_\lambda} e^{i\gamma}$, where $N_1^\lambda$ and $N_2^\lambda$ are
determined purely in terms of $\phi_{NP}$ and $\gamma$. As an
explicit example let us consider NP contributing via the EW penguin to
amplitudes in Eqs.~(\ref{decayamp-2}) and (\ref{decayamp-4}). Using
reparametrization invariance it can easily be absorbed by redefining
the amplitudes $\tilde P^{EW}_{\lambda}$ and $\tilde C_\lambda$, so
that the amplitudes in Eqs.~(\ref{decayamp-2}) and (\ref{decayamp-4})
retain the same form. In general NP contributing to any of the
topological amplitudes can be easily absorbed so that the amplitudes in
Eqs.~(\ref{decayamp-1})--(\ref{decayamp-4}) retain the same form.

The amplitudes for $B\to K^*\rho$ involve three helicities for each of
the modes. These amplitudes and their conjugates, involving the three
helicities, are expressed as
\begin{eqnarray}
\label{eq:fullamps}
 Amp(B\to K^*\rho)&=& \mathcal{A}_0 g_0 + \mathcal{A}_\| g_\| + i
  \, \mathcal{A}_\perp g_\perp~, \nn\\
 Amp({\bar B} \to K^*\rho) &=& \bar {\mathcal{A}}_0 g_0
  + \bar {\mathcal{A}}_\| g_\| - i \, \bar {\mathcal{A}}_\perp g_\perp~,
\end{eqnarray}
where the $g_\lambda$ are the coefficients of the helicity amplitudes written in the
linear polarization basis. The $g_\lambda$ depend only on the angles describing the
kinematics~\cite{Sinha:1997zu}.  The helicity amplitudes (and their conjugate
amplitudes) for the four $K^*\rho$ modes are denoted by $\mathcal{A}_\lambda^{0+}$,
$\mathcal{A}_\lambda^{+-}$, $\mathcal{A}_\lambda^{+0}$,
$\mathcal{A}_\lambda^{00}$, (and $\bar{\mathcal{A}}_\lambda^{0+}$,
$\bar{\mathcal{A}}_\lambda^{+-}$, $\bar{\mathcal{A}}_\lambda^{+0}$,
$\bar{\mathcal{A}}_\lambda^{00}$).
Thus, the number of amplitudes is three times that for the $B \to K \pi$ modes.
In contrast to the $B \to K \pi$ case, one can in principle measure many more
observables in the $B \to K^* \rho$ case. Without including the
interference terms between helicities, one would have three times the number
of observables in comparison to the $K\pi$ modes, {\it i.e.}, 27 observables.
However, many more of observables result from the interference terms
between the helicities.
Let us examine in detail the number of observables available in $B\to K^*\rho$.

The time dependent decay for $B\to f$, where $f$ is one of the $K^*
\rho$ final state, may be expressed as
\begin{eqnarray}
\label{eq:decayrates}
\Gamma(\bbarp(t) \to f) = e^{-\Gamma t} \sum_{\lambda\leq\sigma}
\Bigl(\Lambda_{\lambda\sigma}^f \pm \Sigma_{\lambda\sigma}^f\cos(\Delta M
t) \mp \rho_{\lambda\sigma}^f\sin(\Delta M t)\Bigr) g_\lambda g_\sigma ~,
\end{eqnarray}
where
\begin{eqnarray}
\label{eq:observables_VV}
  B^f_\lambda \equiv \Lambda^f_{\lambda\lambda}=\displaystyle
\frac{1}{2}(|\mathcal{A}^f_\lambda|^2+|\bar \mathcal{A}^f_\lambda|^2),~~&&
\Sigma^f_{\lambda\lambda}=\displaystyle
\frac{1}{2}(|\mathcal{A}^f_\lambda|^2-|\bar \mathcal{A}^f_\lambda|^2),\nn \\[1.ex]
\Lambda^f_{\perp i}= -\!{\rm Im}( \mathcal{A}^f_\perp  \mathcal{A}^{f *}_i \!-\!
\bar \mathcal{A}^f_\perp \bar \mathcal{A}^{f *}_i ),
&&\Lambda^f_{\| 0}= {\rm Re}(\mathcal{A}^f_\| \mathcal{A}^{f *}_0 \!
+\! \bar \mathcal{A}^f _\| {\bar \mathcal{A}^{f *}_0}
), \nn \\[1.ex]
\Sigma^f_{\perp i}= -\!{\rm Im}(\mathcal{A}^f_\perp \mathcal{A}^{f*}_i\!
+\! \bar \mathcal{A}^f_\perp \bar \mathcal{A}^{f *}_i ),
&&\Sigma^f_{\| 0}= {\rm Re}(\mathcal{A}^f_\| \mathcal{A}^{f *}_0\!-\!
\bar \mathcal{A}^{f}_\| \bar \mathcal{A}^{f *}_0
),\nn\\[1.ex]
\rho^f_{\perp i}\!=\! {\rm Re}\!\Bigl(e^{-i\phi^q_{\sss M}} \!\bigl[\mathcal{A}^{f*}_\perp
\bar \mathcal{A}^f_i\! +\! \mathcal{A}^{f *}_i \bar \mathcal{A}^f_\perp\bigr]\Bigr),
&&\rho^f_{\perp \perp}\!=\! {\rm Im}\Bigl(e^{-i\phi^q_{\sss M}}\,
\mathcal{A}^{f *}_\perp
\bar \mathcal{A}^f_\perp\Bigr),\nn\\[1.ex]
\rho^f_{\| 0}\!=\! -{\rm Im}\!\Bigl(e^{-i\phi^q_{\sss M}}[\mathcal{A}^{f*}_\|
\bar \mathcal{A}^f_0\! + \!\mathcal{A}^{f*}_0 \bar \mathcal{A}^f_\| ]\Bigr),
&&\rho^f_{ii}\!=\! -{\rm Im}\!\Bigl(e^{-i\phi^q_{\sss M}} \mathcal{A}^{f*}_i
\bar \mathcal{A}^f_i\Bigr),\nn \\[1.ex]
(\lambda, \sigma = \{ 0, \|, \perp \}, ~ i = \{ 0, \| \}) ~.
\end{eqnarray}
Only for the CP eigenstate $K^{*0}\rho^0$ one can measure all these
18 observables. The other 3 modes are not CP eigenstates so that
time dependent asymmetry cannot be measured. For each of these modes
only $\Lambda_{\lambda\sigma}^f$ and $\Sigma_{\lambda\sigma}^f$ can be
measured, resulting in a total of 12 observables for each of the 3
modes: $B^0\to K^{*+}\rho^-$, $B^+\to K^{*0}\rho^+$ and $B^+\to
K^{*+}\rho^0$. This results in a total of 54 observables.
However, due to the isospin relations in Eqs.~(\ref{eq:isospin-1}) and
(\ref{isospin1}), the number of independent amplitudes is 18.
This results in a total of 35 independent informations related to 18 magnitudes
of the amplitudes and their 17 relative phases at best.
Thus, only 35 of the above 54 observables can be independent.

The modes $B\to K^*\rho$ can be described theoretically using isospin
in a manner analogous to the $B\to K\pi$ modes.  Since there are three
helicity states, the amplitudes corresponding to the different topologies
carry a helicity index and may be denoted by $\tilde T_\lambda$,
$\tilde C_\lambda$, $\tilde A_\lambda$, $\tilde P_\lambda$, $\tilde
P^{ EW}_\lambda$, and $\tilde P^{EW}_{C,\lambda}$. There are hence 18
amplitudes each with its own strong phase denoted by $\dT$,
${\dC}$, ${\dA}$, ${\dP}$, ${\dpew}$ and ${\dpewc}$, respectively.
Since only relative strong phases can be measured, the number of
strong phases may be reduced to 17.  Thus the theoretical description
requires 36 parameters: 18 (real) amplitudes, 17 strong phases and $\gamma$.
Despite the large number of observables in the $K^* \rho$ case, we
still have one more parameter than the observables.

In the next section, we discuss how to determine all the theoretical
parameters, such as the magnitudes and strong phases of the
topological amplitudes, in term of the observables.

\section{Extracting contributions of various topologies}
\label{sec:3}

The $B\to K^*\rho$ modes are described by a total of 36 parameters.
However, as discussed above, one can obtain a maximum
of 35 independent informations from the measurements. Therefore, it
is only possible to solve for the parameters with respect to one
unknown parameter namely $\gamma$. It is well known that the weak
phase $\gamma$ can be measured through certain $B$ decay
processes, such as $B \to D^{(*)} K^{(*)}$~\cite{HFAG,Yao:2006px}.
In this section we present analytic solutions to all the
parameters with respect to $\gamma$. To simplify expression we
introduce some new notation. We define
\begin{eqnarray}
\label{ylambda} y^f_\lambda &=&
\sqrt{1-\Big(\frac{\Sigma^f_{\lambda \lambda}}{\Lambda^f_{\lambda
      \lambda}}\Big)^2}~, \\
\label{phasedef} \alpha_\lambda^{ij} &=&
\arg(\mathcal{A}_\lambda^{ij}) ~, ~~~ \bar \alpha_\lambda^{ij} =
\arg(\bar \mathcal{A}_\lambda^{ij})~,\\
\label{magdef} A^{ij}_\lambda &=& |\mathcal{A}^{ij}_\lambda|~,~~~
\bar A^{ij}_\lambda = |\bar{\mathcal{A}}^{ij}_\lambda|~,
\end{eqnarray}
where $(ij)=(0 +),(+ 0),(+ -), (0 0) $ and $\lambda= \{ 0,\|,\perp
\}$.

For illustration, we divide our task of finding the analytic
solutions into two steps as follows.  We first find the phases
$\alpha^{ij}_{\lambda}$ and $\bar \alpha^{ij}_{\lambda}$ in terms
of observables. Then, using the $\alpha^{ij}_{\lambda}$ and
$\bar \alpha^{ij}_{\lambda}$, we determine the amplitudes
$\tilde{T}_\lambda$, $\tilde{C}_\lambda$, $\tilde{A}_\lambda$,
$\tilde{P}_\lambda$, $\tilde{P}^{EW}_\lambda$,
$\tilde{P}^{EW}_{C,\lambda}$ as well as the strong phases $\delta_{\lambda}^T$,
$\delta_{\lambda}^C$, $\delta_{\lambda}^A$, $\delta_{\lambda}^P$,
$\delta_{\lambda}^{EW}$, and $\delta_{\lambda}^{CEW}$ given in
Eqs.~(\ref{decayamp-1})$-$(\ref{decayamp-4}).

We begin by considering the decay amplitudes of the $K^{*0}\rho^+$ mode
shown in Eq~(\ref{decayamp-1}).
Because the theoretical estimation of the annihilation contribution is very small
$\left( |\tilde{A}/\tilde{P}| \sim \mathcal{O}(\bar \lambda^3) ~{\rm where}~
{\bar \lambda} \sim 0.2  \right)$,
one can safely neglect it~\cite{revtexprob1}.
After neglecting the annihilation terms, we obtain
\begin{eqnarray}
\label{amp0+} \tilde{P}_\lambda &=& A^{0+}_\lambda ~, \\
\label{phs0+} \delta^P_\lambda  &=& \alpha^{0+}_\lambda - \pi ~, \\
\label{ACP0+2} \bar{\alpha}^{0+}_\lambda &=& \alpha^{0+}_\lambda ~.
\end{eqnarray}
These relations imply that the direct CP asymmetry of the $K^{*0} \rho^+$
mode vanishes: $\Sigma^{0+}_{\lambda \lambda}=0 \quad \text{or} \quad
y^{0+}_\lambda =1$.

We set $\alpha^{0+}_0 = \pi ~(\text{or}~\delta^P_0 =0)$ without loss
of generality.  Then the phases $\alpha^{0+}_{\parallel}$ and
$\alpha^{0+}_{\perp}$ can be obtained from the relative phases
$(\alpha^{0+}_\parallel - \alpha^{0+}_0)$ and $(\alpha^{0+}_\perp -
\alpha^{0+}_0)$ that are determined from the angular analysis through
the measurement of $\Lambda^{0+}_{\perp 0}$, $\Sigma^{0+}_{\perp 0}$,
$\Lambda^{0+}_{\parallel 0}$, $\Sigma^{0+}_{\parallel,0}$.
Subsequently all the $\delta^P_\lambda$ and
$\bar{\alpha}^{0+}_\lambda$ for $\lambda = \{0,\parallel,\perp \}$ are
determined from Eqs.~(\ref{phs0+}) and (\ref{ACP0+2}), up to a
discrete ambiguity.  This ambiguity can be removed by using
theoretical estimates~\cite{Aubert:2004xc}. From now on, for the sake
of convenience, we re-parameterize for each helicity state every
relevant phase, such as $\delta^T_{\lambda}$, $\delta^C_{\lambda}$,
$\alpha^{+-}_{\lambda}$, etc., as the relative phase to
$\delta^P_{\lambda}$.  For instance, the strong phase
$\delta^T_{\parallel}$ is understood as $(\delta^T_{\parallel} -
\delta^P_{\parallel})$.

As a next step, we use the isospin analysis to determine $\alpha^{ij}_{\lambda}$,
$\bar \alpha^{ij}_{\lambda}$ in terms of the observables.  The isospin relations
between the decay amplitudes for $B \to K^{*}\rho$ and their conjugate modes
are the same as those given in Eqs.~(\ref{eq:isospin-1}) and (\ref{isospin1}).
Eq.~(\ref{eq:isospin-1}) can be rewritten as
\begin{eqnarray}
  \frac{1}{\sqrt{2}} \Big( A^{0+}_{\lambda} e^{i \alpha^{0+}_{\lambda}}
 -A^{+-}_{\lambda} e^{i \alpha^{+-}_{\lambda}} \Big)
= A^{00}_{\lambda} e^{i \alpha^{00}_{\lambda}}
 - A^{+0}_{\lambda} e^{i \alpha^{+0}_{\lambda}} ~,
\label{eq:isospin-1a}
\end{eqnarray}
where $\mathcal{A}^{ij}_{\lambda} \equiv A^{ij}_{\lambda} e^{i \alpha^{ij}_{\lambda}}$
and $\alpha^{0+}_0 = \pi$ ($\lambda = \{0, \parallel, \perp \}$).
In Eq.~(\ref{eq:isospin-1a}) we note that all the magnitudes $A^{ij}_{\lambda} $
of the decay amplitudes are directly measured and the relative phases
$(\alpha^{ij}_{\parallel} - \alpha^{ij}_{0})$ and
$(\alpha^{ij}_{\perp}- \alpha^{ij}_{0})$
are also measured. Thus, for the three helicity states $\lambda = \{0,\parallel,
\perp \}$, the relevant three isospin relations given in Eq.~(\ref{eq:isospin-1a})
are described by only three independent parameters in total, $\alpha^{+-}_{0}$,
$\alpha^{00}_{0}$ and $\alpha^{+0}_{0}$, which are to be determined. Since we have
3 independent complex equations with these 3 real parameters for $\lambda =
\{0,\parallel, \perp\}$, we can solve these equations to determine the parameters
$\alpha^{0+}_{0}$, $\alpha^{+-}_{0}$ and $\alpha^{+0}_{0}$.
As a result, all the 12 magnitudes $A^{ij}_{\lambda}$ and the 12 phases
$\alpha^{ij}_{\lambda}$ are completely determined.
Details of the solutions of the phases and magnitudes of $\mathcal{A}^{ij}_{\lambda}$
(and $\bar \mathcal{A}^{ij}_{\lambda}$) are given in Appendix \ref{sec:7}.

For the CP conjugate decay modes, one can use the same method as the above by
starting with the isospin relations:
\begin{eqnarray}
  \frac{1}{\sqrt{2}}
  \Big( \bar A^{0+}_{\lambda} e^{i \bar \alpha^{0+}_{\lambda}}
 - \bar A^{+-}_{\lambda} e^{i \bar \alpha^{+-}_{\lambda}} \Big)
= \bar A^{00}_{\lambda} e^{i \bar \alpha^{00}_{\lambda}}
 - \bar A^{+0}_{\lambda} e^{i \bar \alpha^{+0}_{\lambda}} ~,
\label{eq:isospin1a}
\end{eqnarray}
where $\bar \mathcal{A}^{ij}_{\lambda} \equiv A^{ij}_{\lambda}
e^{i \bar \alpha^{ij}_{\lambda}}$ and $\bar \alpha^{0+}_0 = \alpha^{0+}_0 = \pi$
($\lambda = \{0, \parallel, \perp \}$).
Thus, in Eq.~(\ref{eq:isospin1a}) for $\lambda = \{0,\parallel, \perp \}$ there
are only three independent real parameters $\bar \alpha^{+-}_{0}$,
$\bar \alpha^{00}_{0}$ and $\bar \alpha^{+0}_{0}$ that can be determined by solving
the three independent complex equations.  Consequently, all the 12
$\bar \mathcal{A}^{ij}_{\lambda}$ and the 12 $\bar \alpha^{ij}_{\lambda}$ are also
completely determined.

Now let us define the following useful parameters:
\begin{eqnarray}
\label{defX} X_\lambda e^{i \delta^X_\lambda} &=& A_\lambda^{+-}
e^{i \alpha^{+-}_\lambda } - \tilde P_\lambda, \quad \bar
X_\lambda e^{i \bar \delta^X_\lambda } = \bar {A}_\lambda^{+-}
e^{i \bar \alpha^{+-}_\lambda } - \tilde P_\lambda ~, \\
\label{defY} Y_\lambda e^{i \delta^Y_\lambda} &=& \sqrt{2}~
A_\lambda^{00} e^{i \alpha^{00}_\lambda } + \tilde P_\lambda,
\quad \bar Y_\lambda e^{i \bar \delta^Y_\lambda } = \sqrt{2}~ \bar
{A}_\lambda^{00} e^{i \bar \alpha^{00}_\lambda } + \tilde P_\lambda ~.
\end{eqnarray}
Since everything on the right-hand side of these equations has been found,
one can determine for each helicity state all the 8 parameters $X_\lambda,~
\bar X_\lambda,~ Y_\lambda,~ \bar Y_\lambda,~ \delta^X_\lambda,~
\bar \delta^X_\lambda,~ \delta^Y_\lambda,~ \bar \delta^Y_\lambda$ on the left-hand
side in terms of the known parameters by directly solving the 4 complex equations.
Then we reexpress Eqs.~(\ref{decayamp-3}) and (\ref{decayamp-4}) as
\begin{eqnarray}
\label{ADdecayamp-3}
&& X_\lambda e^{i \delta^X_\lambda}
= -e^{i \gamma} \tilde T_\lambda e^{i \delta^T_\lambda}
+ \tilde P^{EW}_{C,\lambda} e^{i \delta^{CEW}_\lambda}~, \\
\label{ADdecayamp-4}
&& Y_\lambda e^{i \delta^Y_\lambda}
= -e^{i \gamma} \tilde C_\lambda e^{i \delta^C_\lambda}
+ \tilde P^{EW}_\lambda e^{i \delta^{EW}_\lambda}~.
\end{eqnarray}
For each $\lambda$, these two complex equations together with their CP conjugate
mode equations ({\it i.e.}, 8 real equations) include 8 real parameters (the
magnitudes and strong phases of 4 topological amplitudes) that need to be determined
in terms of the observables.
It is straightforward to obtain the magnitudes of the topological amplitudes:
\begin{eqnarray}
\label{T} \tilde T_\lambda &= &
 \sqrt{\frac{B^{+-}_\lambda}{2\sin^2\gamma} \Big[
1-{y_\lambda}^{+-}
  \cos (\bar \alpha^{+-}_\lambda -\alpha^{+-}_\lambda)\Big]}~, \\
\label{C} \tilde C_\lambda &= &
 \sqrt{\frac{B^{00}_\lambda}{\sin^2\gamma} \Big[
1-{y_\lambda}^{00}
  \cos \left(\bar \alpha^{00}_\lambda
  -\alpha^{00}_\lambda\right)\Big]}~, \\
\label{CEW} \tilde P^{EW}_{C,\lambda} &=& \sqrt{\frac{1}{4 \sin^2
\gamma} \Big[ X^2_\lambda +\bar X^2_\lambda - 2 X_\lambda \bar
X_\lambda
\cos (\bar \delta^X_\lambda - \delta^X_\lambda + 2 \gamma ) \Big] }~, \\
\label{EW} \tilde P^{EW}_\lambda &=& \sqrt{\frac{1}{4 \sin^2
\gamma} \Big[ Y^2_\lambda +\bar Y^2_\lambda - 2 Y_\lambda \bar
Y_\lambda \cos (\bar \delta^Y_\lambda - \delta^Y_\lambda + 2
\gamma ) \Big] }~.
\end{eqnarray}
And the strong phases are
\begin{eqnarray}
\label{dT} \tan \delta^T_\lambda &=& - \frac{\bar{A}^{+-}_\lambda
\cos \bar \alpha^{+-}_\lambda - A^{+-}_\lambda \cos
\alpha^{+-}_\lambda} {\bar{A}^{+-}_\lambda \sin \bar
\alpha^{+-}_\lambda - A^{+-}_\lambda \sin
\alpha^{+-}_\lambda}~, \\
\label{dC} \tan \delta^C_\lambda &=& - \frac{\bar{A}^{00}_\lambda
\cos \bar \alpha^{00}_\lambda - A^{00}_\lambda \cos
\alpha^{00}_\lambda} {\bar{A}^{00}_\lambda \sin \bar
\alpha^{00}_\lambda - A^{00}_\lambda \sin
\alpha^{00}_\lambda}~, \\
\label{dCEW} \tan \delta^{CEW}_\lambda &=& -
\frac{\bar{A}^{+-}_\lambda \cos (\bar \alpha^{+-}_\lambda +\gamma)
- A^{+-}_\lambda \cos (\alpha^{+-}_\lambda - \gamma) }
{\bar{A}^{+-}_\lambda \sin (\bar \alpha^{+-}_\lambda + \gamma) -
A^{+-}_\lambda \sin (\alpha^{+-}_\lambda - \gamma ) - 2
A^{0+}_\lambda \sin \gamma}~, \\
\label{dEW} \tan \delta^{EW}_\lambda &=& -
\frac{\bar{A}^{00}_\lambda \cos (\bar \alpha^{00}_\lambda +\gamma)
- A^{00}_\lambda \cos (\alpha^{00}_\lambda - \gamma) }
{\bar{A}^{00}_\lambda \sin (\bar \alpha^{00}_\lambda + \gamma) -
A^{00}_\lambda \sin (\alpha^{00}_\lambda - \gamma ) + \sqrt{2}~
A^{0+}_\lambda \sin \gamma}~.
\end{eqnarray}

We have shown that all the hadronic parameters can be cast in terms of the observables
and only one unknown parameter $\gamma$.  If we measure the $\gamma$ from somewhere
else, then we can achieve a model-independent understanding of which hadronic parameter
is dominating in these modes.

Future experiments are important to provide the necessary information to extract each
hadronic parameter.  For instance, the parameters, such as the color-suppressed tree
($\tilde C_{\lambda}$) and the EW penguin ($\tilde P^{EW}_{\lambda}$) amplitudes, can
be determined by using the relevant observables expected to be measured in the near
future and the formulas given in Eqs.~(\ref{C}) and (\ref{EW}).  Then, by comparing
the determined parameters with theoretical predictions, one can further investigate
possible NP effects appearing in $B \to K^* \rho$ decay processes~\cite{London:2004ws}.
In the next section we discuss in details how the determination of these parameters
in terms of the observables can be used to verify the hierarchy relations between the
topological amplitudes that are conventionally assumed to be true in the SM.  We also
discuss ways to test the validity of assumptions equating the strong phases of a
certain set of topological amplitudes.

\section{Testing the hierarchy of topological amplitudes and possible
  relations between their strong phases}
\label{sec:4}

In last section we have estimated all the topological amplitudes and strong phases
purely in terms of the observables and $\gamma$. Having obtained these relations it
is straightforward to conclude that if there exist any relations between the
theoretical parameters they must also result in relations among the observables.

We first derive certain relations between the observables that test the conventional
hierarchy between the topological amplitudes within the SM.  It may be expected that
\begin{eqnarray}
  \label{eq:topoheir}
  \dsp\frac{\tilde T_\lambda}{\tilde P_\lambda}\approx \frac{\tilde
    P^{EW}_\lambda}{\tilde P_\lambda} \approx \bar{\lambda}~,~~~~
  \dsp\frac{\tilde C_\lambda}{\tilde P_\lambda}\approx\frac{\tilde
    P^{EW}_{C,\lambda}}{\tilde P_\lambda}\approx \bar{\lambda}^2 ~~
\end{eqnarray}
in analogy to the expectations~\cite{GHLR} for the modes $B\to K\pi$, as the two modes
are topologically equivalent.

The topological amplitudes $\tilde P_\lambda$, $\tilde A_\lambda$, $\tilde T_\lambda$,
$\tilde C_\lambda$, $\tilde P^{EW}_\lambda$ and $\tilde P^{EW}_{C\lambda}$ have been
expressed in terms of the observables and $\gamma$ in the previous section. It is
therefore easy to see that there must exist a relation between the observables and
$\gamma$ that must hold as a consequence of the hierarchy between the topological
amplitudes.  The relations (\ref{eq:topoheir}) indicate the hierarchy relation
$\tilde P _\lambda >  \tilde T_\lambda \approx \tilde P^{EW}_\lambda >
\tilde C_\lambda \approx \tilde P^{EW}_{C  \lambda}$ which must hold within the SM.

A simple approach would be to test the hierarchy $ \tilde P_\lambda > \tilde T_\lambda
> \tilde C_\lambda   $, which would imply the following relation:
\begin{eqnarray}
& & 2 \sin^2 \gamma B^{0+}_\lambda > B^{+-}_\lambda
\Big[1-{y_\lambda}^{+-}
  \cos (\bar \alpha^{+-}_\lambda - \alpha^{+-}_\lambda )\Big]
   > 2 B^{00}_\lambda \Big[ 1-{y_\lambda}^{00}
  \cos (\bar \alpha^{00}_\lambda -\alpha^{00}_\lambda ) \Big]~.
\label{TEST1}
\end{eqnarray}
To test the hierarchy $\tilde P^{EW}_\lambda > \tilde P^{EW}_{C,\lambda}$, one can
test the following relation:
\begin{eqnarray}
\label{TEST2} Y^2_\lambda + \bar Y^2_\lambda - 2 Y_\lambda \bar Y_\lambda
\cos (\bar \delta^Y_\lambda - \delta^Y_\lambda + 2 \gamma)
> X^2_\lambda + \bar X^2_\lambda - 2 X_\lambda \bar X_\lambda
\cos (\bar \delta^X_\lambda - \delta^X_\lambda + 2 \gamma)~.
\end{eqnarray}
Besides the above relation, simple tests verifying the hierarchy of
$\tilde P^{EW}_\lambda$ and $\tilde P^{EW}_{C\lambda}$ can be derived. Assuming that
$\tilde P^{EW}_{C\lambda} = \bar{\lambda}^2 ~\tilde P_\lambda$ in
Eq.~(\ref{decayamp-3}), it can be shown that
\begin{equation}
  \label{eq:pewc-1}
  \frac{{B^{+-}_\lambda}\Big[1-{y_\lambda}^{+-}
    \cos(\bar \alpha^{+-}_\lambda-\alpha^{+-}_\lambda+2\gamma)\Big]}
    {2 \sin^2 \gamma B^{0+}_\lambda}
    =1+{\cal O} ({\bar\lambda}^2 )~
\end{equation}
Similarly assuming that $P^{EW}_{\lambda}=\bar{\lambda}~ P_\lambda$ in
Eq.~(\ref{decayamp-4}), it can be found that
\begin{equation}
  \label{eq:pew-1}
  \frac{{B^{00}_\lambda}\Big[1-{y_\lambda}^{00}
    \cos(\bar \alpha^{00}_\lambda-\alpha^{00}_\lambda+2\gamma)\Big]}
    {\sin^2 \gamma B^{0+}_\lambda}
    =1+{\cal O} (\bar\lambda )~
\end{equation}
The relations $\tilde T_\lambda \approx \tilde P^{EW}_\lambda$ and
$\tilde C_\lambda \approx \tilde P^{EW}_{C  \lambda}$ would imply that
\begin{eqnarray}
\label{TEST3} 2 B^{+-}_\lambda \Big[1-{y_\lambda}^{+-}
  \cos (\bar \alpha^{+-}_\lambda - \alpha^{+-}_\lambda )\Big]
\approx Y^2_\lambda + \bar Y^2_\lambda - 2 Y_\lambda \bar
Y_\lambda \cos (\bar \delta^Y_\lambda - \delta^Y_\lambda + 2 \gamma)~, \\
\label{TEST4} 4 B^{00}_\lambda \Big[1-{y_\lambda}^{00}
  \cos (\bar \alpha^{00}_\lambda - \alpha^{00}_\lambda )\Big]
\approx X^2_\lambda + \bar X^2_\lambda - 2 X_\lambda \bar
X_\lambda \cos (\bar \delta^X_\lambda - \delta^X_\lambda + 2 \gamma)~.
\end{eqnarray}
Testing the hierarchy $ ~\tilde T_\lambda > \tilde P^{EW}_{C,\lambda}$ and
$\tilde P^{EW}_\lambda > \tilde C_\lambda$ is rather simple.  We note that
$\tilde T_\lambda$ and $ \tilde C_\lambda$ can be rewritten as
\begin{eqnarray}
\label{T-2} \tilde T_\lambda &=& \sqrt{\frac{1}{4 \sin^2 \gamma}
\Big[ X^2_\lambda +\bar X^2_\lambda - 2 X_\lambda \bar X_\lambda
\cos (\bar \delta^X_\lambda - \delta^X_\lambda) \Big] }~, \\
\label{C-2} \tilde C_\lambda &=& \sqrt{\frac{1}{4 \sin^2 \gamma}
\Big[ Y^2_\lambda +\bar Y^2_\lambda - 2 Y_\lambda \bar Y_\lambda
\cos (\bar \delta^Y_\lambda - \delta^Y_\lambda ) \Big] }~.
\end{eqnarray}
Comparing these equations with Eqs.~(\ref{CEW}) and (\ref{EW}), we find that
\begin{eqnarray}
\label{TEST6} \tilde T_\lambda > \tilde P^{EW}_{C,\lambda} \Longrightarrow
&& \sin \left(\bar \delta^X - \delta^X + \gamma \right) < 0 ~, \\
\label{TEST7} \tilde P^{EW}_\lambda > \tilde C_\lambda \Longrightarrow
&& \sin \left(\bar \delta^Y - \delta^Y + \gamma \right) > 0 ~.
\end{eqnarray}

The relations (\ref{TEST1})$-$(\ref{TEST7}) will provide a litmus test to verify the
hierarchy assumption between the magnitudes of topological amplitudes.  Since the
decay modes $B \to K \pi$ and $B \to K^* \rho$ are equivalent at quark level, the
same hierarchy relation is expected to hold in $B \to K \pi$ modes.

Now let us move our focus on to the strong phases.  One can derive
several relations purely in terms of the observables and $\gamma$ by
assuming relations between the strong phases of the topological
amplitudes. These relations would be very important in verifying the
assumptions often made between these strong phases, such as $\dC
\approx \delta^{CEW}_\lambda \approx \dP$ and $\dpew \approx \dT$,
which are expected to hold within the SM. We discuss only a few such
relations that test some common assumptions being made on the strong
phases~\cite{revtexprob2}.

We first consider the implication of the interesting relation $\dC = \dP$.  In our
convention ($\dP \equiv 0$), it means $\dC = 0$. So from Eq.~(\ref{dC}) we get the
relation
\begin{eqnarray}
\label{TEST8} \bar A^{00}_\lambda \cos \bar \alpha^{00}_\lambda =
A^{00}_\lambda \cos \alpha^{00}_\lambda ~.
\end{eqnarray}
In fact, this relation can be obtained directly from Eq.~(\ref{decayamp-4}).
If $\dC = 0$, the real part of the amplitude $\mathcal{A}^{00}_\lambda$ is the same
as that of its CP conjugate amplitude $\bar \mathcal{A}^{00}_\lambda$, which is
just the restatement of the relation (\ref{TEST8}).
Since the topological amplitude $\tilde C_\lambda$ is estimated to be very small
($\tilde C_{\lambda} = \bar \lambda^2 \tilde P_{\lambda}$) in
the SM, it is also expected from Eq.~(\ref{decayamp-4}) that in the SM the direct
CP asymmetry in the $K^{*0} \rho^0$ mode almost vanishes.
Similarly the assumption $\delta^{CEW}_\lambda = \dP$ leads to the relation
\begin{eqnarray}
\label{TEST9}\bar A^{+-}_\lambda \cos (\bar \alpha^{+-}_\lambda
+\gamma) = A^{+-}_\lambda \cos (\alpha^{+-}_\lambda - \gamma)~,
\end{eqnarray}
obtained from Eq.~(\ref{dCEW}).
Finally, from the assumption $\dT = \dpew$, we get the relation
\begin{eqnarray}
\label{TEST10}
- \frac{\bar{A}^{+-}_\lambda \cos \bar \alpha^{+-}_\lambda
- A^{+-}_\lambda \cos \alpha^{+-}_\lambda}
{\bar{A}^{+-}_\lambda \sin \bar \alpha^{+-}_\lambda
- A^{+-}_\lambda \sin \alpha^{+-}_\lambda}
= - \frac{\bar A^{00}_{\lambda} \sin \bar \alpha^{00}_{\lambda}
+ A^{00}_{\lambda} \sin \alpha^{00}_{\lambda}}
{ 2 A^{00}_{\lambda} \cos \alpha^{00}_{\lambda}
+ \sqrt{2} A^{0+}_{\lambda}}~ [ 1 + {\cal O} (\bar \lambda^2) ]~,
\end{eqnarray}
where we have used $\tilde C_{\lambda} = \bar \lambda^2 \tilde P_{\lambda}$.

Before concluding this section we note that the validity of the
several relations derived above, or the degree to which they fail to
hold, will shed light on the possible origins of the ``$B\to K\pi$
puzzle,'' and hence help in uncovering possible NP contributions.

\section{Isolating signals of new physics in $B \to K^* \rho$ modes }
\label{sec:4}

In the previous section we derived relations that test possible
relations between topological amplitudes and strong phases. If we
instead assume that these relations hold within the SM, a violation of
these relations would signal NP. In Sec.~\ref{sec:2} we showed that
NP contributing to any of the topological amplitudes can be easily
absorbed so that the amplitudes in Eqs.~(\ref{decayamp-1})$-$(\ref{decayamp-4})
retain the same form, making it impossible to
have a clean signal of NP. However, if there exist relations between
the amplitudes or strong phases, the $B\to K^*\rho$ amplitudes would
differ from the SM form in the presence of NP.  Since the number of
independent SM parameters is reduced, one may now, not only be able to
see signals of NP but also solve for NP parameters. In this section we
consider two cases to explore this possibility.  We first discuss the
consequence of relations between $\tilde P^{EW}_\lambda e^{i\dpew}$
and $\tilde T_\lambda e^{i\dT}$, and $\tilde P^{EW}_{C,\lambda}
e^{i\dpewc}$ and $\tilde C_\lambda e^{i\dC}$ that are expected to hold
in the SM. We next consider the case where the strong phases are
related in the SM.

Let us consider NP contributing via the EW penguins, to amplitudes in
Eqs.~(\ref{decayamp-2}) and (\ref{decayamp-4}). We assume that NP
contributes with an amplitude $N_\lambda e^{i\delta^N_\lambda}
e^{i\phi_{NP}}\equiv N_1^\lambda e^{i\delta^N_\lambda}+N_2^\lambda
e^{i\delta^N_\lambda} e^{i\gamma}$. We have shown that using
angular-analysis we can solve for all the parameters in
Eqs.~(\ref{decayamp-1})$-$(\ref{decayamp-4}). In particular we can
measure $\tilde P^{EW}_\lambda$, $\tilde P^{EW}_{C,\lambda}$, $\tilde
T_\lambda$, $\tilde C_\lambda$, $\dpew$, $\dpewc$, $\dT$, $\dC$ in
terms of $\gamma$. Note that these measured values include any NP
contributions that may be present.  In fact, if NP contributes via the
EW penguins, the SM amplitudes (defined by calligraphic
characters)-- $\tilde {\cal P}^{EW}_{\lambda}$ and $\tilde {\cal
  C}_\lambda$ are the only ones modified by NP, but they cannot
themselves be measured. The other amplitudes are unmodified by NP and
hence we need not distinguish the SM amplitudes from the amplitudes
defined in Eqs.~(\ref{decayamp-1})$-$(\ref{decayamp-4}).

In the SM, to a good approximation~\cite{EWPs}, using flavor SU(3) the
$\Delta I=3/2$ parts of the tree and electroweak penguin Hamiltonians
are simply related by
\begin{equation}
  \label{eq:Hamiltionian}
  {\cal H}^{EW}_{\Delta I=3/2} = -\frac32  \left[ {c_9 +
      c_{10} \over c_1 + c_2} \right] \left\vert {V_{tb}^* V_{ts}
      \over V_{ub}^* V_{us}} \right\vert
  {\cal H}^{tree}_{\Delta I=3/2}~.
\end{equation}
Hence, $\tilde {\cal P}^{EW}_\lambda e^{i\dpew}$ and $\tilde
P^{EW}_{C,\lambda} e^{i\dpewc}$ are related to $\tilde T_\lambda
e^{i\dT}$ and $\tilde {\cal C}_\lambda e^{i\dC}$ :
\begin{eqnarray}
  && \tilde {\cal P}^{EW}_\lambda e^{i\dpew} \simeq {3\over 2} \left[ {c_9 +
      c_{10} \over c_1 + c_2} \right] \left\vert {V_{tb}^* V_{ts} \over V_{ub}^* V_{us}}
\right\vert ~\tilde T_\lambda
  e^{i\dT}\equiv \zeta\tilde T_\lambda
  e^{i\dT} ,~~ \nn\\
  && \tilde P^{EW}_{C,\lambda} e^{i\dpewc} \simeq {3\over 2} \left[ {c_9
      + c_{10} \over c_1 + c_2} \right] \left\vert {V_{tb}^* V_{ts} \over V_{ub}^* V_{us}}
\right\vert ~\tilde {\cal C}_\lambda
  e^{i\dC}\equiv\zeta\tilde {\cal C}_\lambda e^{i\dC}~,
\label{diagrels}
\end{eqnarray}
where the $c_i$ are Wilson coefficients~\cite{BuraseffH}.
Using the SM relations in Eq.~(\ref{diagrels}), it is easy to see that
\begin{eqnarray}
  \zeta \tilde C_\lambda e^{i\dC}&=& \zeta \tilde {\cal C}_\lambda
  e^{i\dC}+ \zeta N_2^\lambda  e^{i\delta^N_\lambda}~,\nn\\
  \label{eq:relation-1}
  &=& \tilde P^{EW}_{C,\lambda}
  e^{i\dpewc}+\zeta N_2^\lambda  e^{i\delta^N_\lambda}~,\\
  \tilde P^{EW}_\lambda e^{i\dpew}&=& \tilde {\cal P}^{EW}_\lambda
  e^{i\dpew}+N_1^\lambda  e^{i\delta^N_\lambda}~,\nn\\
  \label{eq:relation-2}
  &=&\zeta\tilde T_\lambda e^{i\dT}+N_1^\lambda  e^{i\delta^N_\lambda}~.
\end{eqnarray}
Eqs.~(\ref{eq:relation-1}) and (\ref{eq:relation-2}) form four
relations in terms of only three unknowns $N_1^\lambda$, $N_2^\lambda$
and $\delta^N_\lambda$ for each $\lambda$. Hence $N_1^\lambda$,
$N_2^\lambda$ and $\delta^N_\lambda$ can easily be solved. $\phi_{NP}$
the weak phase of NP can be obtained from the relation
\begin{equation}
  \label{eq:phi-gamma}
  \frac{N_1^\lambda}{N_2^\lambda}=
  \frac{\sin(\gamma-\phi_{NP})}{\sin\phi_{NP}}~.
\end{equation}
In fact, under the assumptions made there are enough observables even
to solve for $\zeta$, enabling us not only to measure NP but also test
the SU(3) assumption.

Instead of assuming that the SM amplitudes are related by
Eq.~(\ref{diagrels}), we next assume that in the SM the strong phases
are related such that $\dC=\dpewc=\dP$ and $\dpew=\dT$.  It is then
easy to conclude that
\begin{eqnarray}
  \label{eq:pew-pewsm-1}
  \tilde P^{EW}_\lambda e^{i\dpew} &=& \tilde {\cal P}^{EW}_{\lambda}
  e^{i\dT}+N_1^\lambda e^{i\delta^N_\lambda}\\
  \label{eq:C-Csm-1}
  \tilde C_\lambda e^{i\dC} &=& \tilde {\cal C}_\lambda e^{i\dP}+
  N_2^\lambda  e^{i\delta^N_\lambda}
\end{eqnarray}
We again have four equations, but now in terms of five unknowns
$\tilde {\cal P}^{EW}_{\lambda}$, $\tilde {\cal C}_\lambda$,
$N_1^\lambda$, $N_2^\lambda$ and $\delta^N_\lambda$. Hence if the
strong phases are related in the SM, the failure of the relations
$\dC=\dpewc=\dP$ and $\dpew=\dT$ can be tested, but it is not possible
to solve for any of the NP parameter.  Hence, the assumptions on
strong phases do not allow a clean test of NP; the failure of the
relations between strong phases could be due to NP or simply due to
hadronic effects within the SM. The assumptions of Eq.~(\ref{diagrels})
in the case discussed previously allow a clean test of NP as the
assumption can be verified independent of possible NP contributions.


We emphasize that due to reparameterization invariance it is in
general impossible to have a clean signal (i.e. model-independent
signal) of NP~\cite{Imbeault:2006nx}. Furthermore, tests of NP are
based on relations between the amplitudes and strong phases of
topological amplitudes. It is not always possible to independently
test the hadronic assumption and at the same time cleanly measure the
NP parameters. However, we do demonstrate that if the tree and
color-supressed tree are related to the electroweak penguins and
color-supressed electroweak penguins by the well known relations of
Eq.~(\ref{diagrels}), {\em it is possible not only to verify the
  validity of these relations but also to have a clean measurement of
  New Physics parameters}. It would be worth doing an angular anlysis
in $B\to K^*\rho$ not only to establish cleanly the validity of the
relations in Eq.~(\ref{diagrels}) but also at the same time to cleanly
probe for NP.

\section{Numerical Analysis}
\label{sec:4}

In this section we perform numerical analysis to investigate how much sensitive
to possible NP effects each observable for $B \to K^* \rho$ decays could be.
As discussed in Sec.~V, we consider NP contributing via the EW penguins.
For simplicity we further assume that additional information on the theoretical
parameters is given from somewhere, for instance, from future theoretical
estimates, so that the SM amplitudes $\tilde {\cal C}_\lambda$ and
$\tilde {\cal P}^{EW}_{\lambda}$ are known.
Therefore, the SM amplitude $\tilde {\cal P}^{EW}_{\lambda}$
is the only one modified by NP and the amplitudes $\tilde C_{\lambda}$ and
$\tilde P^{EW}_{\lambda}$ in Eqs.~(\ref{decayamp-2}) and (\ref{decayamp-4}) can
be expressed explicitly in terms of the SM and NP amplitudes:
\begin{eqnarray}
 \tilde C_\lambda &=& \tilde {\cal C}_\lambda ~, \\
 \tilde P^{EW}_\lambda e^{i\dpew} &=& \tilde {\cal P}^{EW}_\lambda e^{i\dpew}
 +\tilde P^{~ \prime EW}_\lambda e^{i\delta^{~ \prime EW}_\lambda} e^{i\phi_{EW}}~,
\label{eq:relation-3}
\end{eqnarray}
where $\delta^{~ \prime EW}_\lambda$ and $\phi_{EW}$ are the strong and weak
phases of the NP amplitudes, respectively.

\begin{table}
\caption{Experimental data on the CP-averaged branching ratios ($\cal B$ in units of
$10^{-6}$), the longitudinal polarization fractions ($f_L$), and the direct CP asymmetries
(${\cal A}_{CP}$) for $B \to K^* \rho$ modes~\cite{HFAG}.}
\smallskip
\begin{tabular}{|c|c|c|c|}
\hline
  Mode & ${\cal B}~(10^{-6})$ & $f_L$ & ${\cal A}_{CP}$   \\
\hline
$B^+ \to K^{*0} \rho^+$ & ~~~$9.2 \pm 1.5$~~~ & ~~~~$0.48 \pm 0.08$~~~ & $-0.01 \pm 0.16$ \\
$B^+ \to K^{*+} \rho^0$ & $3.6 \pm 1.9$ & $0.96^{+0.06}_{-0.16}$ & $0.20^{+0.32}_{-0.29}$  \\
$B^0 \to K^{*+} \rho^-$ & $5.4 \pm 3.9$ & $-$ & $-$  \\
$B^+ \to K^{*0} \rho^0$ & $5.6 \pm 1.6$ & $0.57 \pm 0.12$ & $0.09 \pm 0.19$  \\
\hline
\end{tabular}
\label{table:1}
\end{table}

We first summarize the present status of the experimental results on $B \to K^* \rho$ modes
in Table~\ref{table:1}~\cite{HFAG}.  So far only two modes $B^+ \to K^{*0} \rho^+$ and
$B^+ \to K^{*+} \rho^0$ have been observed.
For illustration, we perform a numerical study to make predictions on the physical observables.
Because those data shown in Table~\ref{table:1} are currently the only available ones, we follow
the following procedures: (i) In order to determine the theoretical parameters in a reasonable
way, we adopt the $\chi^2$ minimization technique and use the currently available data as
constraints on the parameters.  We first consider only the dominant strong penguin contributions
and neglect all the other topological amplitudes.  Then, we use five known observables given
in Table~\ref{table:1} and try to fit the dominant strong penguin amplitudes $\tilde P_{\lambda}$
$(\lambda = 0, \perp, \parallel)$ and their phases $\delta^P_{\lambda}$ with $\delta^P_0 \equiv 0$.
The degrees of freedom ({\it d.o.f}) for this fit is 0.
As a next step, we assume that the SM amplitudes, such as $\tilde T$, $\tilde {\cal C}$,
$\tilde {\cal P}^{EW}$, $\tilde A$, follow the conventional hierarchy as in $B \to K \pi$ within
the SM: for instance, in the pQCD approach~\cite{Li:2005kt}, ${\tilde T} / {\tilde P} =0.15$,
${\tilde {\cal P}^{EW}} / {\tilde P} =0.12$, ${\tilde {\cal C}} / {\tilde P} =0.04$,
${\tilde A} / {\tilde P} =0.005$.  Their phases are set to be small.
(ii) Using the parameters determined in the previous step, we calculate all the 35 observables
within the SM.  We use $\gamma = 62^{\circ}$.
(iii) To investigate the possible NP effects, we consider two different cases:
a case with sizable but relatively small NP effects and another case with relatively large NP
effects.  For the former case, we assume $r_{_{EW}} = 0.12$ and $r_{_{EW}}^{\prime} = 0.05$,
while for the latter, we assume $r_{_{EW}} = 0.12$ and $r_{_{EW}}^{\prime} = 0.20$,
where $r_{_{EW}} \equiv \tilde {\cal P}^{EW} / \tilde P$ and
$r_{_{EW}}^{\prime} \equiv \tilde P^{\prime~ EW} / \tilde P$.
In both cases, the strong phases are set to be $\delta^{\prime~ EW}_{\lambda}
= \delta^{EW}_{\lambda}$.  The NP weak phase is chosen to be $\phi_{EW} =90^{\circ}$, which
is consistent with that used to explain the $B \to K \pi$ puzzle~\cite{KOY}.

\begin{figure}[h]
\centerline{\epsfig{figure=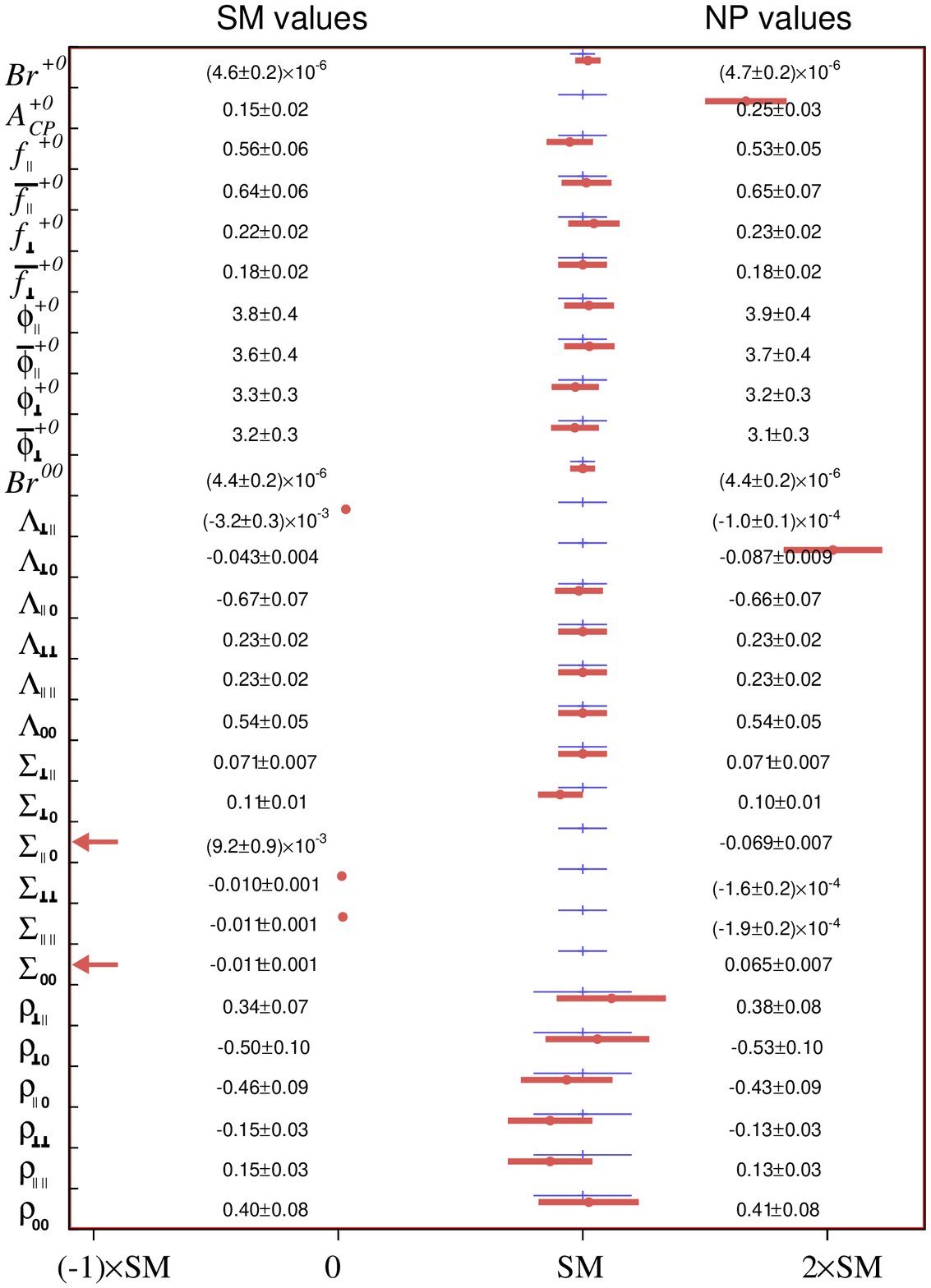,angle=0, scale=0.75} }\
 \caption{ The case of $r_{_{EW}} = 0.12$ and $r_{_{EW}}^{\prime} = 0.05$,
 where $r_{_{EW}} \equiv \tilde {\cal P}^{EW} / \tilde P$ and
 $r_{_{EW}}^{\prime} \equiv \tilde P^{\prime~ EW} / \tilde P$.
 The SM values are shown as the ``thin'' bars and the numbers
 in the left column.  The predictions with the NP contributions are shown
 as the ``thick'' bars and the numbers in the right column.}
\label{fig:1NP}
\end{figure}

\begin{figure}[h]
\centerline{\epsfig{figure=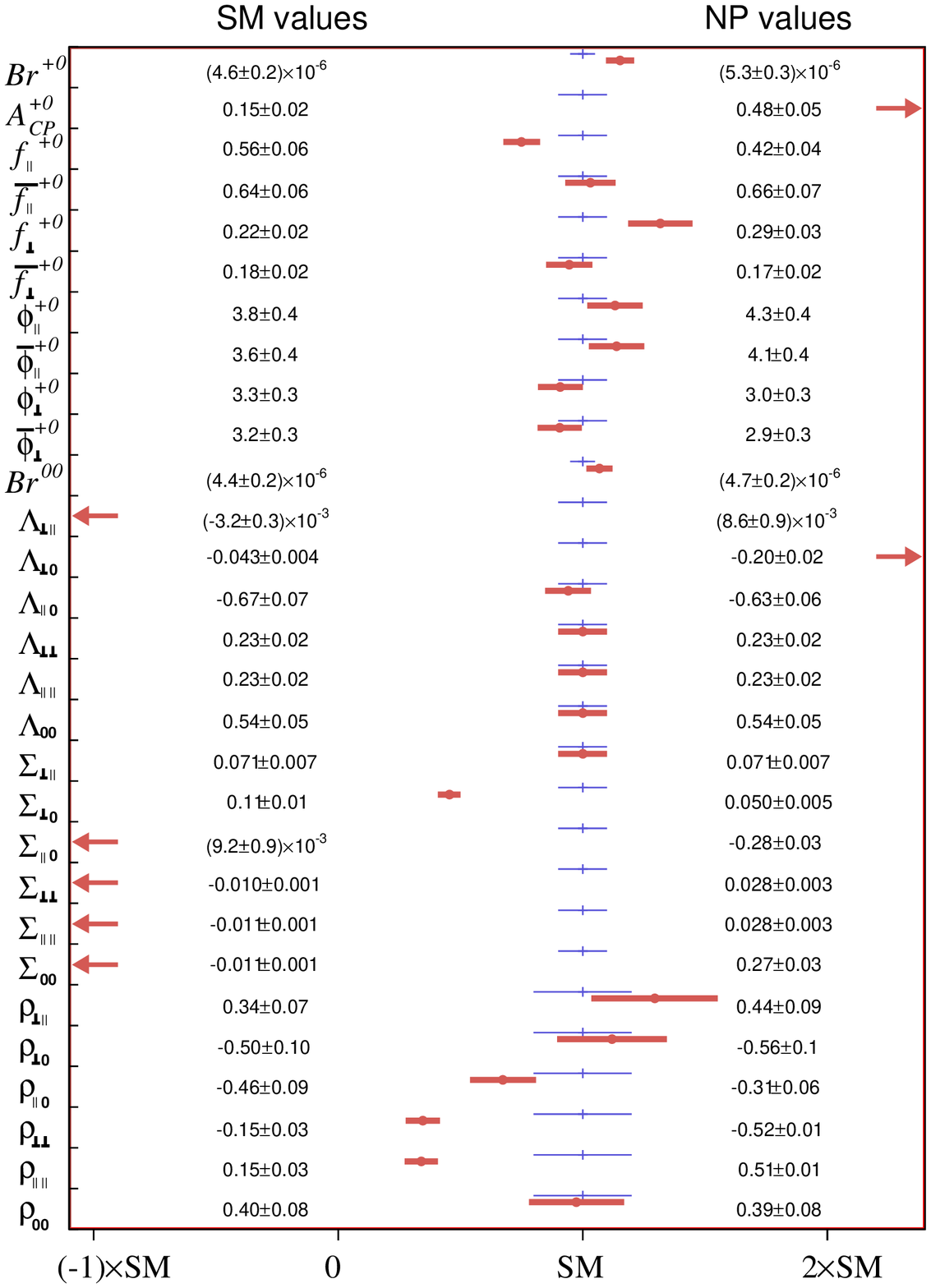,angle=0, scale=0.75} }
 \caption{ The case of $r_{_{EW}} = 0.12$ and $r_{_{EW}}^{\prime} = 0.20$,
 where $r_{_{EW}} \equiv \tilde {\cal P}^{EW} / \tilde P$ and
 $r_{_{EW}}^{\prime} \equiv \tilde P^{\prime~ EW} / \tilde P$.
 The SM values are shown as the ``thin'' bars and the numbers
 in the left column.  The predictions with the NP contributions are shown
 as the ``thick'' bars and the numbers in the right column.}
\label{fig:2NP}
\end{figure}

The results are presented in Figs.~\ref{fig:1NP} and \ref{fig:2NP}.
Here the observables $\phi^{ij}_{\parallel, ~\perp}$ and ${\bar \phi}^{ij}_{\parallel, ~\perp}$
defined as the relative phases ${\rm arg} \left( A^{ij}_{\parallel, ~\perp} / A^{ij}_{0} \right)$
and ${\rm arg} \left( {\bar A}^{ij}_{\parallel, ~\perp} / {\bar A}^{ij}_{0} \right)$ respectively,
are included, since they have been measured through $B^0 \to \phi K^{*0}$~\cite{Aubert:2004xc}.
The uncertainties depicted in
the figures have been assumed just for illustration as follows.  The BRs are expected to be
measured accurately so that their errors have been assumed to be 5\%.  The other observables
including the direct CP asymmetries, the polarization fractions, etc, except the observables
$\rho_{ii}$, have been assumed to be measured with 10\% errors.  The uncertainties in
$\rho_{ii}$ have been assumed to be 20\%.

In Fig.~\ref{fig:1NP}, the physical observables are predicted
within the SM as well as with the NP effects of
$r_{_{EW}}^{\prime} = 0.05$. Here the notation $Br^{ij}$ denotes
the branching ratio for $B \to K^{*i} \rho^j$, and others are
similarly defined.  The SM values are shown as the ``thin''
bars and the numbers in the left column.
The predictions with the NP contributions are shown as the ``thick''
bars and the numbers in the right column. It is
clearly shown that certain observables are very sensitive to the
NP effects in $B \to K^* \rho$ decays: for instance,
$A_{CP}^{+0}$, $\Lambda_{\perp \parallel}$, $\Lambda_{\perp 0}$,
$\Sigma_{\perp \perp}$, $\Sigma_{\parallel \parallel}$,
$\Sigma_{00}$, and $\Sigma_{\parallel 0}$. In particular, it is
interesting to note that the direct CP asymmetry for the mode $B^+
\to K^{*+} \rho^0$, which has been already observed but with large
errors yet, is sensitive to the NP effect. Fig.~\ref{fig:2NP}
shows predictions with the NP effects of $r_{_{EW}}^{\prime} =
0.20$. Here the SM values are the same as those in
Fig.~\ref{fig:1NP}.  We see that for many observables, the
predictions with the NP contributions are very much off the SM
ones.  This is the expected result, due to the large NP effects.
Thus, if any anomalously large NP effects, e.g., a large
color-suppressed tree contribution, appears in $B \to K^* \rho$
decays, one can easily find them through those observables.
Obviously if more (precise) experimental data are available in the
future, all the predictions shown in Fig.~\ref{fig:1NP} and
~\ref{fig:2NP} will be able to become more reliable.

\section{Conclusions}
\label{sec:5}

We have performed a detailed study of the $B \to K^* \rho$ decays
using a model-independent approach. It was shown that $B\to K^*\rho$
modes have a distinct advantage due the large number of independent
observables that can be measured. In comparison to the $B \to K \pi$
modes that yield only 9 independent observables, the $B\to K^*\rho$
modes result in as many as 35 independent observables. Since $B\to
K\pi$ and $B\to K^*\rho$ have the same quark level subprocess, the
study of $B\to K^*\rho$ may well shed light on the well known ``$B\to
K\pi$ puzzle.''  The relevant decay amplitudes were decomposed into
linear combinations of the topological amplitudes with their
respective strong phases assuming isospin.  We point out that the
amplitude written this way are the most general ones and included
contributions not only from the SM but also any NP that might exsit.
We obtain explicit model-independent expressions for all the
topological amplitudes and their strong phases in terms of observables
and the weak phase $\gamma$. With $\gamma$ measured using other modes,
our results are the first in literature to estimate the topological
amplitudes and strong phases purely in terms of observables, for the
$B\to K\pi$ analogous modes. We further suggest clean tests to verify
if there exist any hierarchy relations among topological amplitudes
analogous to the ones conventionally assumed to exists for $B\to K\pi$
in the SM.  In addition we present tests that would verify any
equality between the strong phases of the topological amplitudes. A
model independent understanding of the relative sizes of the
topological amplitudes and relations between their strong phases could
provide valuable insights into NP searches.  While it is not in
general possible to independently test the hadronic assumption and at
the same time cleanly measure the NP parameters, we show one example
where it is possible to do both.  We demonstrate that if the tree and
color-supressed tree are related to the electroweak penguins and
color-supressed electroweak penguins, it is not only possible to
  verify the validity of such relations but also to cleanly measure
  New Physics parameters.  We also present a numerical study to
examine which of the observables are more sensitive to New Physics.

\vspace{1cm}
\centerline{\bf ACKNOWLEDGMENTS}
\noindent
The work of C.S.K. was supported in part by CHEP-SRC and in part by the KRF
Grant funded by the Korean Government (MOEHRD) No. KRF-2005-070-C00030.
The work of S.O. was supported by the Second Stage of Brain Korea 21 Project.
The work of Y.W.Y. was supported by the KRF Grant funded by the Korean Government
(MOEHRD) No. KRF-2005-070-C00030.


\newpage
\appendix

\section{ Determination of  $ \mathcal{A}_\lambda^f $  and  $
  \bar{\mathcal{A}}_\lambda^{f} $ with observables}
\label{sec:7}

\subsection{Determination of the magnitude $ \mathcal{A}^f_\lambda$ and
 $\bar{\mathcal{A}}^f_\lambda$}

The branching ratios (BRs) and direct CP asymmetries of the decay modes
$B \to K^* \rho$ are measured experimentally~\cite{HFAG,Yao:2006px}. Using the
measured values of BRs and direct CP asymmetry of each helicity for the decay
modes, $A^{f}_\lambda ~(\equiv |\mathcal{A}^f_\lambda|~)~$ can be determined
straightforwardly. The direct CP asymmetry is defined as $a^{f}_\lambda \equiv
\frac{\Sigma^f_{\lambda\lambda}}{B^f_\lambda} $, where
$\Sigma^f_{\lambda\lambda}$ and $B^f_\lambda$ are defined in
Eq.~(\ref{eq:observables_VV}), and $f$ is one of the final states of $K^* \rho$.
Therefore, $A^{f}_\lambda $ and $\bar A^{f}_\lambda $ can be written as
\begin{eqnarray}
(A^{f}_\lambda)^2 & = & B^f_\lambda + \Sigma^f_{\lambda\lambda}~~~~\mathrm{and}
~~~~(\bar{A}^{f}_\lambda)^2 = B^f_\lambda - \Sigma^f_{\lambda\lambda}~.
\end{eqnarray}

\subsection{Determination of the phases of $ \mathcal{A}^{f} _\lambda$  and
$\bar{\mathcal{A}^{f}_\lambda}$ }

Let us first try to find out the phases $\alpha^{ij}_{\lambda}$ of
$\mathcal{A}^{f}_\lambda$.  Since the relative phases $(\alpha^{ij}_{\parallel}
-\alpha^{ij}_0 )$ and $(\alpha^{ij}_{\perp} -\alpha^{ij}_0 )$ can be measured in
experiment, one needs to determine only $\alpha^{ij}_0$.
We express the three equations of Eq.~(\ref{eq:isospin-1a}) explicitly with three
unknown parameters $\alpha^{+-}_0,~ \alpha^{00}_0$, and $\alpha^{+0}_0$:
\begin{eqnarray}
\label{isospin-eq}
\frac{1}{\sqrt{2}} \Big( A^{0+}_0 e^{i \pi} -A^{+-}_0 e^{i \alpha^{+-}_0} \Big)
 &=& A^{00}_0 e^{i \alpha^{00}_0} - A^{+0}_0 e^{i \alpha^{+0}_0} ~, \nn\\
\frac{1}{\sqrt{2}} \Big[ A^{0+}_\parallel e^{i \pi}
 -A^{+-}_\parallel e^{i ( \alpha^{+-}_0 + \tilde \phi^{+-}_\parallel )} \Big]
&=& A^{00}_\parallel e^{i (\alpha^{00}_0 + \tilde \phi^{00}_\parallel )}
 - A^{+0}_\parallel e^{i (\alpha^{+0}_0 + \tilde \phi^{+0}_\parallel ) } ~, \nn\\
\frac{1}{\sqrt{2}} \Big[ A^{0+}_\perp e^{i \pi}
 -A^{+-}_\perp e^{i (\alpha^{+-}_0 +\tilde \phi^{+-}_\perp )} \Big]
&=& A^{00}_\perp e^{i (\alpha^{00}_0 + \tilde \phi^{00}_\perp )}
 - A^{+0}_\perp e^{i (\alpha^{+0}_0 + \tilde \phi^{+0}_\perp )} ~,
\label{iso-eq}
\end{eqnarray}
where $\tilde \phi^{ij}_\parallel$ and $\tilde \phi^{ij}_\perp$ are defined in
terms of the observables $\phi^{ij}_\parallel$ $(\equiv \alpha^{ij}_\parallel
-\alpha^{ij}_0)$ and $\phi^{ij}_\perp$ $(\equiv \alpha^{ij}_\perp -\alpha^{ij}_0)$
such that
\begin{eqnarray}
\label{tilde-phi}
\tilde{\phi}^{ij}_\parallel = \phi^{ij}_\parallel-\phi^{0+}_\parallel
~~~ {\rm and}~~~
\tilde{\phi}^{ij}_\perp = \phi^{ij}_\perp-\phi^{0+}_\perp ~.
\end{eqnarray}
Here we remind that in our convention each phase $\alpha^{ij}_{\parallel (\perp)}$
has been defined as the relative phase to $\delta^P_{\parallel (\perp)}
=\alpha^{0+}_{\parallel (\perp)} -\alpha^{0+}_0 \equiv \phi^{0+}_{\parallel (\perp)}$.
Then we can re-write Eq.~(\ref{isospin-eq}) as the matrix equation
\begin{equation}
\label{iso-mat} \mathbf{S} \mathbf{X} = \mathbf{A} ~,
\end{equation}
where the matrix $\mathbf{S}$ and the column vectors $\mathbf{X}$ and $\mathbf{A}$
are given by
\begin{eqnarray}
\label{matS}
\mathbf{S} =\left(
\begin{array}{cccccc}
\frac{1}{\sqrt{2}}A^{+-}_0  & 0 &
                  A^{00}_0  & 0 &
                 -A^{+0}_0  & 0 \\
0 & \frac{1}{\sqrt{2}}A^{+-}_0  &
                  0 & A^{00}_0  &
                  0 &-A^{+0}_0  \\
 \frac{1}{\sqrt{2}}A^{+-}_\parallel \cos \tilde
\phi^{+-}_\parallel  &  -\frac{1}{\sqrt{2}}A^{+-}_\parallel
\sin\tilde \phi^{+-}_\parallel & A^{00}_\parallel\cos\tilde
\phi^{00}_\parallel &  -A^{00}_\parallel\sin \tilde
\phi^{00}_\parallel &  -A^{+0}_\parallel\cos \tilde
\phi^{+0}_\parallel &  A^{+0}_\parallel\sin \tilde \phi^{+0}_\parallel \\%
 \frac{1}{\sqrt{2}}A^{+-}_\parallel \sin \tilde
\phi^{+-}_\parallel &  \frac{1}{\sqrt{2}}A^{+-}_\parallel \cos
\tilde \phi^{+-}_\parallel  & A^{00}_\parallel\sin \tilde
\phi^{00}_\parallel & A^{00}_\parallel\cos \tilde
\phi^{00}_\parallel & -A^{+0}_\parallel\sin \tilde
\phi^{+0}_\parallel & -A^{+0}_\parallel\cos \tilde
\phi^{+0}_\parallel \\
 \frac{1}{\sqrt{2}}A^{+-}_\perp \cos \tilde
\phi^{+-}_\perp & -\frac{1}{\sqrt{2}}A^{+-}_\perp \sin \tilde
\phi^{+-}_\perp  & A^{00}_\perp\cos \tilde \phi^{00}_\perp &
-A^{00}_\perp\sin \tilde \phi^{00}_\perp & -A^{+0}_\perp\cos
\tilde \phi^{+0}_\perp & A^{+0}_\perp\sin \tilde \phi^{+0}_\perp \\%
\frac{1}{\sqrt{2}}A^{+-}_\perp \sin \tilde \phi^{+-}_\perp &
\frac{1}{\sqrt{2}}A^{+-}_\perp \cos \tilde \phi^{+-}_\perp &
A^{00}_\perp\sin \tilde \phi^{00}_\perp & A^{00}_\perp\cos\tilde
\phi^{00}_\perp & -A^{+0}_\perp\sin \tilde \phi^{+0}_\perp &
-A^{+0}_\perp\cos \tilde \phi^{+0}_\perp
\end{array} \right)  \nn \\
\label{vecX}
\mathbf{X} = \left(~~
\begin{array}{cccccc}
\cos \alpha^{+-}_0 ~& \sin \alpha^{+-}_0 ~& \cos \alpha^{00}_0 ~&
\sin \alpha^{00}_0 ~& \cos \alpha^{+0}_0 ~& \sin \alpha^{+0}_0
\end{array} \right)^T ~, ~~~~~~~~~~~~~~~~~~~~~~~~~~~~ \nn \\
\mathbf{A} = \left(~~
\begin{array}{cccccc}
-\frac{1}{\sqrt{2}} A^{0+}_0 ~&~ 0 ~&~ -\frac{1}{\sqrt{2}}
A^{0+}_\parallel ~&~ 0 ~&~ -\frac{1}{\sqrt{2}} A^{0+}_\perp ~&~
0~~
\end{array} \right)^T ~, ~~~~~~~~~~~~~~~~~~~~~~~~~~~~~~~~~~~~
\end{eqnarray}
where $\mathbf{X}$ is the column vector to be determined.
One can easily solve this matrix equation by calculating the inverse matrix of
$\mathbf{S}$. The solution $\mathbf{X}$ is given by
\begin{equation}
\label{alpha-sol} \mathbf{X} = \mathbf{S}^{-1} \mathbf{A} ~.
\end{equation}
We note that in Eq.~(\ref{alpha-sol}) both cosine and sine of each phase $\alpha^{+0}_0,
\alpha^{00}_0, \alpha^{0+}_0$ can be determined, which results in removing discrete
ambiguities associated with trigonometrical functions of the solution.

By using exactly the same method as above, one can also find the phases
$\bar \alpha^{ij}_{\lambda}$ of the CP conjugate amplitudes
$\bar{\mathcal{A}^{f}_\lambda}$.

\begin{thebibliography}{99}
\bibitem{revtexprob1} In this case, the number of parameters is
still the same as that of informations.  As 6 parameters ($\tilde A_\lambda$,
$\delta^A_\lambda$) are neglected, the number of informations are also reduced
by 6 ($i.e.$, $\bar A^{0+}_\lambda$, $\bar \alpha^{0+}_\lambda$) due to the
vanishing direct CP asymmetry of the $K^{*0} \rho^+$ mode.
\bibitem{revtexprob2} More relations between observables under various
  assumptions on strong phases are also available.  Furthermore,
  though the results shown in Secs.~III and IV have been obtained by
  neglecting the annihilation contribution, more general results
  including the annihilation contribution are also available in
  detail.  Interested readers may contact one of the authors.
\end{thebibliography}
\end{document}